# Enhancing the Efficiency in Checking Constraints Satisfaction when Planning Ground-based and Space Experiments, Using an Alternative Problem

Atanas Marinov Atanassov

*Solar-Terrestrial Influences Laboratory, Department in Stara Zagora*

E-mail: At_M_Atanassov@yahoo.com

**Summary**
 The situational analysis lies in the basis of space and ground-based experiment planning. It is connected with the use of complex computation models of environment and with verification of the restricting conditions, due to the character of the conducted experiments and the solved scientific tasks.
 The present work proposes a formulation of the situational analysis on the basis of the finite abstract automata theory. On this basis, optimization of the situational analysis is suggested by formal schemes for adaptation to the conditions of the model environment. The efficiency enhancement is illustrated by results from the application of the proposed real-time optimization for photometric system control.

**Introduction:** To solve scientific and practical problems, connected with ground-based and satellite experiments and measurements, special activity is required– planning, connected with checking the satisfaction of a multitude of geometrical, physical and other constraints [1]. The planning will involve the application of an adequate mathematical model for analysis of a number of constraints and for their formulation, this representing the so-called situational analysis (SA). It allows to determine suitable time intervals to conduct experiments and measurements and to optimize complex and expensive scientific programmes. The practice of the situational analysis involves check-up of different constraints, such as:

– a satellite crosses the Earth's shadow;
– A satellite is within the Earth's radiation belt zone, within the zone of the shock wave or of the magnetopause of the Earth's magnetosphere [2];
– the angle between the observed object and another bright object on the celestial sphere (Sun, Moon, bright starts) is less than a definite value [3];

- a satellite passes over a territory of the Earth's surface [4];
- a satellite is within the visibility zone of an observation point on the Earth's surface;
- The Earth's radiation background is within admissible limits with a view to the conducted measurements [3];
- A satellite is appropriately orientated towards the force lines of the Earth's magnetic field;
- The magnetic force line where the measuring device is located, pierces the polar oval region;
- The visibility axis of the optical instrument falls within a region of the celestial sphere or the Earth's surface, which is interesting from the point of view of the conducted experiment.

The appropriate orientation of the model of the space platform with scientific equipment on board in respect to the simulation of experiments and measurements in the objective space will be called a situation. Each situation can be presented in the general case by a predicate function $S$:

(1) $\quad S = S(\vec{R}, A, t) = \overline{0,1}$

In (1) $\{\vec{R}\} = \langle \vec{r}_1(t), \vec{r}_2(t), ..., \vec{r}_n(t) \rangle$ - the multitude of the radii-vectors of the objects in the model space, $\{A\} = \langle \alpha_1(t), \alpha_2(t), ..., \alpha_m(t) \rangle$ - the multitude of vector or scalar fields, describing certain properties of the model space – and $t$ – the time. Due to the complex character of the processes and phenomena, occurring in the space, $S$ may have a complex analytical representation.

Actually, we can have a combination of several constraints restricting conditions. In addition, we shall examine such conditions that are independent on one another (none of them is represented by the others). Thus, the multitude of conditions $\{\gamma_i\}$ can be juxtaposed to the multitude of predicate functions $\{s_i\}$.

The implementation of situation $s$ will require the fulfillment of the following identity:

(2) $\quad S = s_1 \wedge s_2 \wedge ... \wedge s_n = 1$

**Presenting the planning process by finite automata formalism**

It is obvious that the check-up of the identity of (2) requires calculation of the predicate functions $s_i$, which is connected with considerable calculation

(expenses). According to (2) we can consider $\{s_i\}$ as being arranged in the sense of consequences (computation sequence) of calculation of $s_i$:

(3)  $S = (...(s_1 \wedge s_2) \wedge ... \wedge s_{n-1}) \wedge s_n$

In (3), every predicate function $s_i$ is calculated if those before it in $\{s_i\}$ have value of one. A measure for the ineffectiveness of scheme (3) can be the expenses for calculation of $s_i$ for i<k, when $s_k = 0$.

Every discrete-deterministic model may be treated as abstract finite automata [6] and presented by 6-tuple:

(4)  $F = <Z, X, Y, \varphi, \psi, z_0>$,

functioning in discrete automata time. In (4), Z is the set of internal states, X is the set of incoming (input) signals, Y is set of exit (output) signals, $z(t+\Delta t) = \varphi(z(t), x(t))$ – transition function and $z_0$ - initial state. For the classical Mealy's automata we have:

(5)  $\begin{cases} z(t+\Delta t) = \varphi(z(t), x(t)) \\ y(t) = \phi(z(t), x(t)) \end{cases}$

We can apply this formal approach to the situation analysis. For moment $t$, the set of input signals $X(t)$ is the total of all values determining the discrete model. In many cases, these are vectors which determine the measurement place or the observation direction. The internal states of the situational analysis automata (SAA) can be characterized by information entropy, which is proportional to the quantity of processed information for one cycle, related to moment t. These states are related with the implementation of different model calculations, corresponding to the different constraints (situational conditions) $\gamma_i$ whose satisfaction is verified for every cycle of automata. We can reason that the state $z(t)$ will depend on the order of situational conditions $\gamma_i$ in $\{\gamma_i\}$, and on the evolution of the mathematical model. With applying (3), each state of the automata will depend on the input signal $x(t)$ only and will not depend on the previous state of the automata:

(6)  $z(t+\Delta t) = \varphi(x(t))$

Automata without storage (such as the SAA) have only one stable state, which coincides with the initial one. Here we have an example of such automata which may be excited in terms of one cycle consecutively to different internal states. According to (5), for the SAA to function, no storage is necessary. This leads to simplicity and easy accomplishment. However, the lack of a storage and an appropriate transition function makes SAA ineffective for space research applications.

It is important to have in mind that each constraint $\gamma_i$ is executed in a time interval $T_i$ and is not executed in the next one $T_i^*$. The only practically imposed requirements for $T_i$ and $T_i^*$ that we shall take into account are that they are finite and $T_i, T_i^* \gg \Delta t$.

A priori information about the feasibility execution of conditions $\gamma_i$ with time is laching. The check-up of $\gamma_i$ can be verified by tracing the evolution of the deterministic numerical model, underlying (1).

**Adaptation and Optimization Strategy**

Problem A will be called inverse to B, if the objectives of A and B are contrary [7]. In our case the problem for specifying the temporal interval, in which the conditions $\{\gamma_i\}$ are satisfied, is inverse to the problem for specifying its adjacent in which they are not satisfied or, according to (3) at least one of them is not satisfied. Obviously, this problem is more efficient since it is sufficient to check whether only one condition $\gamma_i$ is not satisfied.

$$(7) \quad \overline{S} = \overline{s_1 \wedge s_2 \wedge ... \wedge s_k \wedge ... \wedge s_n} = \overline{s_1} \vee \overline{s_2} \vee ... \vee \overline{s_k} \vee ... \vee \overline{s_n}$$

In fact, according to the above-said, $T_i \gg \Delta t$. If, for discretization moment $t$ we have $s_k = 0$, then we can presume that for $t + \Delta t$ this same condition will be satisfied again.

The strategy for implementing situational analysis is reduced to a consecutive verification of the condition. If all conditions are satisfied, then we have a situation of the searched type. In the case when for moment t a condition is found out which is not satisfied, there is not a situation. When for $t'$ this condition is dissatisfied, once again the next condition in the set $\{\gamma_i\}$ is being verified.

Here, we shall examine two approaches, related with the organisation of the conditions' verifications. The first approach is related with replacement of the first encountered condition in the first place of the set of conditions [8]. Instead of (7), as a result of the disjunction commutation, we can write down:

$$(8)\ \overline{S} = \overline{s}_k \vee \overline{s}_1 \vee \overline{s}_2 \vee ... \vee \overline{s}_{k-1} \vee \overline{s}_{k+1}...\overline{s}_n$$

The verifying algorithm for (3) always begins from the first element. The replacement of each dissatisfied condition when applying the alternative strategy is equivalent to adaptation of (3) to the conditions of the model medium. The latter is a function of the transition $\psi(z(t), x(t))$.

The second approach requires to treat the set of conditions as a ring-shaped structure [9]. This means that the last element of $\{\gamma_i\}$ is followed by the first one:

$$(9)\ \overline{S} = \overline{s}_k \vee \overline{s}_{k+1} \vee ... \vee \overline{s}_n \vee \overline{s}_1 \vee ... \vee \overline{s}_{k-1}$$

This allows to apply the alternative strategy without rearranging the set of conditions only by moving a pointer along this ring-shaped structure until finding a dissatisfied condition. The shifting of the pointer along the ring-shaped structure until finding dissatisfied condition plays the role of a transition function. A situation is found out when a full circle along the ring-shaped structure is completed.

**Analysis and Effectiveness Estimation**

In order to estimate the effectiveness, the following two cases are of interest:

a). The predicate functions $s_i$ on the basis of which identify (2) is verified, which are characterized by equal or almost equal computational expenses $\xi_i \cong const$;

b). The predicate functions $s_i$ on the basis of which identify (2) is verified, which are characterized by different computational expenses.

In the first case, for each step of SAA performance based on the offered strategy, $(k-1).\xi$ computational units will be saved in contrast to the direct application of (3) where the dissatisfied condition is in the $k^{th}$ position. The transition function will not be of any significance since the different predicate functions are characterised by equal computational expenses.

As an illustration we will give an analysis, related with the transition of a couple of satellites over the visibility zone of a ground-based observation station, proper mutual configuration and distance between them, lack of factors hampering the observation of satellites from the Earth (Moon), lack of factors, preventing the observation of the optical instruments form one of the satellites (Moon, Sun). This situation considers the problem for a possible synchronous observation form the Earth's surface and from one of the satellites of an emission, caused by neutral particles injection for studying the Eearth's ionosphere (the *Aktiven* experiment).

In the second case where the computational expenses for the separate conditions are different, for the saved computational expenses we can write down $\sum_{i=1}^{k-1} \xi_i$. Without going into detail, it is worth noting that the arrangement of the conditions in $\{\gamma_i\}$ (3) and the transition function $z(t)$ are significant. Despite the quasicausal choice of an unfeasible condition in applying the alternative strategy, there are certain peculiarities. The first algorithm may carry a condition with large computational expenses to the first position. If this condition remains in the beginning of the set of conditions $\{\gamma_i\}$, the effectiveness of the analysis may be reduced. With the second approach such a problem does not exist as a result of the constant arrangement of the conditions and their circulation.

We have an example of a situational problem of the second type when including in the analysis the conditions, related with the parameters of the medium $\alpha_i$. The condition for the angle between the satellite velocity vector and the magnetic field vector $\vec{B}(\vec{r},t)$ to have an appropriate value is important for the analysis of active experiments. (The Aktiven experiment). The radiation background is important in astrophysical observations of roentgen sources [3]. When phenomena connected with particles dissipation in the polar oval region are investigated, it is important to measure the magnetic field in the region of the force lines along which the dissipation occurs [6]. In all presented cases the verified conditions are related with complex model computations.

**Example for application**

For effective real time control of a zenith photometer [11] it is required to check conditions, related with the Sun's and Moon's position for elimination of their influence on the measurements. The restriction for the Sun is to be at an angle φ´ below the horizon and for the Moon - to close with direction to the zenith an angle bigger than φ´´.

Two check-up conditions contain two versions for arranging $\{\gamma_i\}$. In the first one, the verification of the condition for the Sun is in the first place and for the Moon- in the second; in the other version the places are reverse. To perform the analysis for one year period with a time step of one minute without optimization, 412,264 checks for the Sun's position and 525,600 for the Moon's position are made, respectively, by the first arrangement version, 525,600 and 171,025 checks are performed by the second version. The application of the optimization approach leads to solving the same task with 468,161 and 186,659 checks only.

**Conclusion**

This paper examines the situational analysis of space experiments on the basis of the finite automata theory. It allows formalization of the computing processes and search of possibilities for optimization of the computing algorithms at a higher abstraction level.

The advantages of the suggested approach are obvious even in the examined simple application case. The properties of the examined optimization can be revealed best in the analysis with participation of a larger number of check-up conditions, which, on the other hand, are connected with considerable computational expenses. During real-time experiment planning and, especially during their control, the analysis' acceleration can be of major importance.

Besides, a strategy is examined for optimization of the situational analysis, based on the solution of the inverse problem. This, on the other hand, allows to apply algorithms for adaptation to the conditions of the model environment. Two algorithms are suggested whose application can considerably enhance the analysis' efficiency.